\documentclass[12pt]{article}
\usepackage{amsmath,amssymb,bm,mathrsfs}
\usepackage[utf8x]{inputenc}
\usepackage[T1]{fontenc}
\usepackage{graphicx}
\begin{document} 
\begin{titlepage}
\begin{center}
\begin{large}
{\bf Inflationary models that generalize the constant roll constraint}

\end{large}
\vskip0.5truecm

B. Boisseau\footnote{E-mail: Bruno.Boisseau@lmpt.univ-tours.fr} and H. Giacomini\footnote{E-mail: Hector.Giacomini@lmpt.univ-tours.fr}

\vskip0.5truecm
{\sl CNRS--Institut Denis Poisson (UMR7013)\\Université de Tours, Université d'Orléans\\Parc Grandmont, 37200 Tours, France}

\vskip0.2truecm

\end{center}
\vskip1truecm
\begin{abstract}
Recently a class of inflationary models satisfying the constant rate of roll constraint, 
$\ddot{\phi}+(3+\alpha)H\dot{\phi}=0$, has been studied and compared with the latest cosmological observational data. We consider the broader class of  constraints $ \ddot{\phi}+\frac{\alpha_1}{2}\dot{\phi}+\alpha_2 H\dot{\phi}=0$ and  find an exact particular  solution, without initial singularity, which is an attractor of the dynamics governed by a positive periodic potential $V(\phi)=E+ F \cos (\Omega  \phi )+G \cos (2 \Omega  \phi )$.  The spectral index $n_s$ and the tensor to scalar ratio $r$ are consistent with the recent observational data. 
\end{abstract}

\end{titlepage}

\section{Introduction}\label{introduction}

Inflation is considered as the best proposal to solve the horizon and the flatness problems in cosmology~\cite{starobinsky, guth, linde, AAPJS}. It makes predictions about present properties of the universe which have been confirmed by numerous experimental observations. However there are many possible models of inflation and at the present time, the choice is, still, very broad. The slow-roll models with a single scalar field $\phi$ which predict gaussianity at a good level of precision are the topic  of considerable scrutiny~\cite{weinberg, peter-uzan,martin_infl}. 

One way to ensure that  inflation is sufficient to describe our Universe, is to choose the potential $V(\phi)$ ``sufficiently flat''. If 
$V(\phi)$ is constant, the Klein-Gordon equation for $\phi$, is reduced to 

$$\ddot{\phi}(t)+3H(t)\dot{\phi}(t)=0.$$

This model, named  ultra-slow-roll-inflation, was considered in 
~\cite{woodard,kinney}. It  manifestly violates one of the conditions of the slow-roll model. 

More recently a new class of models appeared: the constant-roll models, introduced in ~\cite{JMHMTS, HMAASJY} and further studied in 
~\cite{HMAAS, yi-gong,odintsov_oikonomou,dimopoulos,gao,Karam:2017rpw}. This class of models is characterized by the condition of ``constant rate of roll'':

$$\ddot{\phi}(t)+(3+\alpha)H(t)\dot{\phi}(t)=0,$$

which generalizes the ultra-slow-roll inflation. Here $\alpha$ is an arbitrary parameter.
 
 It is shown in \cite{HMAASJY} that it is possible, under this condition, to find particular exact solutions of the inflationary model,  associated to a definite simple potential $V(\phi)$.

These solutions are very useful for understanding  the salient properties of these models. It is shown in \cite{HMAAS} that a model, which  can be rendered compatible with the latest observational results on the spectral index $n_s$ can be found; in particular, such a model admits an explicit solution, which is an attractor for the inflationary dynamics.

In this paper we show that the efficient method~\cite{HMAASJY,HMAAS} allows imposing a  more general condition:
	
		 $$ \ddot{\phi}(t)+\left(\frac{\alpha_1}{2}+\alpha_2 H(t)\right)\dot{\phi}(t)=0,$$
		where $\alpha_1$ and $\alpha_2$ are arbitrary parameters. When $\alpha_1=0$ and $\alpha_2=3+\alpha$ we recover the condition employed in ~\cite{JMHMTS, HMAASJY}.
		This condition can also be written as:
		$$\dot{H}(t)+\alpha_1 H(t)+\alpha_2 H^2(t)=\alpha_3,$$
		where $\alpha_3$ is a new arbitrary parameter. As  will be shown in section 2, by differentiating  this last equation and using  the equations of motion, we obtain the preceding condition.    
		
		This condition  yields a model of inflation with a periodic potential of a simple form:
		
		$$V(\phi)=E+ F \cos (\Omega  \phi )+G \cos (2 \Omega  \phi ),$$
		
		which is new, to our knowledge. The method of determining this potential from the new imposed condition enables us to find an exact particular solution of the equations of motion, which is, also,  an attractor of the dynamics. 
		
		The quantities $E, F, G, \Omega$ are functions of the three arbitrary parameters $\alpha_1$, $\alpha_2$ and $\alpha_3$. The explicit expressions are given in section 3. This new condition is a simple mathematical generalisation of the condition employed in  ~\cite{JMHMTS, HMAASJY}. 
		
		With the new term that we have added, the equation for $H(t)$ becomes a complete Riccati equation with constant coefficients. 
		
		While there is no physical motivation for this generalisation, {\em a priori}, it turns out to have interesting physical implications.
		The goal of the present work is to describe  the physical implications of this mathematical generalisation. We have found, indeed,  that the addition of the new parameter $\alpha_1$ has very important  consequences for  the physical interpretation of the model.

Our model does describe an inflationary phase, without an initial singularity, but it predicts, also, after a deceleration period, a new phase of accelerated expansion. In this article we take the first phase of acceleration as the inflationary period and we will not analyse the second phase of acceleration.

In section 2 we show that there exist two potentials that are compatible with the new condition imposed on the scalar field and we determine also exact particular solutions of the equations of motion, associated with these potentials.

In section 3 we focus on  only one of the two potentials because the other one, in spite of the presence of three arbitrary parameters, is not compatible with the recent experimental results. We analyse the stability of the associated particular solution and we find the condition that ensures that this solution is an attractor of the dynamics. 
We determine also explicit expressions of the time $t_1$ where inflation ends and the time $t_2$ where a new acceleration begins. 

In section 4 we study the scalar perturbations and their associated power spectrum and, from this calculation, determine the spectral index $n_s$. 

In section 5 we find that the values of the parameters of the model can be chosen in such a way that the spectral index $n_s$ is consistent with the latest experimental values. 

In section 6 the tensor perturbations are studied and the tensor to scalar ratio $r$ is calculated. The obtained values are found to be compatible with the recent  observational data~\cite{planck}.  
 
 In section 7 the study of the super-Hubble evolution of the curvature perturbations shows that there is no growing mode;  only a constant mode survives. 
 
 In section 8 we present our conclusions.

\section{ Models of inflation that generalize the constant roll condition}

Let us consider the action of a single scalar field minimally coupled to gravity  modeling an inflationary scenario:

\begin{equation}
\label{inflationaction}
S = \int d^4x \sqrt{-g}\left(\frac{1}{2\kappa^2}R-\frac{1}{2}g^{\mu\nu}\partial_{\mu}\phi\partial_{\nu}\phi-V(\phi)\right).
\end{equation}
In this work we use the following system of units and constants:

$\hbar=c=1,$                         

$\kappa^2=8\pi G=\frac{1}{M_{Pl}^2},$

$G=2.905\times10^{-87}\mathrm{s}^{2}$  ,  \; $\kappa=2.70205\times10^{-43}\mathrm{s}.$

In a flat FRLW metric $ds^2=-dt^2+a(t)d{\bf x}^2$ we obtain the Friedmann equations and the equation of motion for the scalar field:

\begin{equation}
\label{eqa}
3H^2(t)=\kappa^2\left(V\left(\phi(t)\right) +\frac{1}{2}\dot{\phi}^2(t)\right),
\end{equation}

\begin{equation}
\label{eqb}
\dot{H}(t)=-\frac{\kappa^2}{2}\dot{\phi}^2(t),
\end{equation}

\begin{equation}
\label{eqc}
\ddot{\phi}(t)+3H(t)\dot{\phi}(t)+V_{,\phi}\left(\phi(t)\right)=0,
\end{equation}

where $H(t)=\frac{\dot{a}(t)}{a(t)}$ is the Hubble parameter.

We search for an analytic particular solution of these cosmological equations, that satisfies the following generalization of the constant-roll constraint:   
\begin{equation}
\label{eqe}
\ddot{\phi}(t)+\left(\frac{\alpha_1}{2}+\alpha_2 H(t)\right)\dot{\phi}(t)=0. 
\end{equation}
This supplementary condition determines a particular class of potential $V(\phi)$.
It is easy to see that this constraint can be obtained from the equation
\begin{equation}
\label{eqd}
\dot{H}(t)+\alpha_1 H(t)+\alpha_2 H^2(t)=\alpha_3, 
\end{equation}
where $ \alpha_1,\alpha_2,\alpha_3 $ are arbitrary real constants.
Indeed, taking the derivative of this equation (\ref{eqd}) and using the equation (\ref{eqb})  we arrive to the  constraint 
(\ref{eqe}).
We prefer to use the constraint in this last form (\ref{eqd}), which is a Riccati equation with constant coefficients and can be explicitly integrated. We see that the new term that we have added in (\ref{eqe}) gives a complete Riccati equation with constant coefficients for $H(t)$.
The general solution of this equation can be expressed in two equivalent forms: 

\begin{equation}
\label{H_a}
H(t)=\frac{\omega\coth(\frac{1}{2}\omega(t-\tilde{C_1}))-\alpha_1}{2\alpha_2},
\end{equation} 

and
\begin{equation}
\label{H_b}
H(t)=\frac{\omega\tanh(\frac{1}{2}\omega(t-C_1))-\alpha_1}{2\alpha_2},
\end{equation} 
% Which are equivalent if the integration constants $C_1$ are complex but are different if  $C_1$ are real
where $\tilde{C_1}$ and $C_1$ are  arbitrary integration constants and 
$$\omega=\sqrt{\alpha_1^2+4\alpha_2\alpha_3}.$$

We are interested in real solutions with $\tilde{C_1}$, $C_1$ and $\omega$ real parameters. In this case  (\ref{H_a}) and 
(\ref{H_b}) are different. From equation (\ref{eqb}) we deduce that these two solutions must be decreasing, so for (\ref{H_a}), $\alpha_2$ must be positive and for (\ref{H_b}) $\alpha_2$ must be negative.

Introducing the derivative of the solutions (\ref{H_a}) and (\ref{H_b}) in (\ref{eqb}) we obtain   differential equations of first order in $\phi(t)$ which can be integrated and give, respectively:
\begin{equation}
\label{Phi_a}
\phi(t)=\tilde{C_2}-\frac{\sqrt{2}\ln[\tanh\left(\frac{1}{4}\omega\left(t-\tilde{C_1}\right)\right)]}{\kappa\sqrt{\alpha_2}},
\end{equation} 
and

\begin{equation}
\label{Phi_b}
\phi(t)=C_2+\frac{2\sqrt{2}\arctan[\tanh\left(\frac{1}{4}\omega\left(t-C_1\right)\right)]}{\kappa\sqrt{-\alpha_2}}.
\end{equation} 

We determine now the potentials corresponding to these solutions.
From (\ref{eqa}) we have
\begin{equation}
\label{eqaa}
V\left(\phi\right)=\frac{3}{\kappa^2}H^2(t) -\frac{1}{2}\dot{\phi}^2(t).
\end{equation}
We introduce $H(t)$ from (\ref{H_a}) and $\dot{\phi}^2(t)$ from (\ref{eqb}) in (\ref{eqaa}) and we obtain:
\begin{equation}
\label{V_a(t)}
V\left(\phi\left(t\right)\right)=\frac{3\left(-\alpha_1+\omega\coth(\frac{1}{2}\omega(t-\tilde{C_1}))\right)^2}{4\alpha_2^2\kappa^2}-
\frac{\omega^2}{4\alpha_2\kappa^2}\frac{1}{\sinh^2(\frac{1}{2}\omega(t-\tilde{C_1)}}.
\end{equation}

This expression gives the potential $V$ in terms of $t$ but we want to obtain the potential in function of the scalar field $\phi$. In order to eliminate $t$ we employ the following identity, that we obtain from (\ref{Phi_a}):
$$\tanh\left(\frac{1}{4}\omega\left(t-\tilde{C_1}\right)\right)=
\exp\left(-\frac{\kappa\sqrt{\alpha_2}}{\sqrt{2}}\left(\phi-\tilde{C_2}\right)\right)$$
and using the expression of $\coth\left(\frac{1}{2}\omega\left(t-\tilde{C_1}\right)\right)$ and 
$\sinh\left(\frac{1}{2}\omega\left(t-\tilde{C_1}\right)\right)$ as functions of \linebreak $\tanh\left(\frac{1}{4}\omega\left(t-\tilde{C_1}\right)\right)$ we finally obtain for the potential
\begin{equation}
\label{V_a}
V\left(\phi\right)=\frac{3(-\alpha_1+\omega\cosh[\frac{\sqrt{\alpha_2}\kappa(\phi-\tilde{C_2})}{\sqrt{2}}])^2}{4\alpha_2^2\kappa^2}- 
\frac{\omega^2\left(-1+\cosh[\sqrt{2}\sqrt{\alpha_2}\kappa(\phi-\tilde{C_2})]\right)}{8\alpha_2\kappa^2}.
\end{equation}
For the second model  we introduce $H(t)$ from (\ref{H_b}) and $\dot{\phi}^2(t)$ from (\ref{eqb}) in (\ref{eqaa}) and we arrive to the following expression for the potential in terms of $t$:
\begin{equation}
\label{V_b(t)}
V\left(\phi\left(t\right)\right)=\frac{3\left(-\alpha_1+\omega\tanh(\frac{1}{2}\omega(t-C_1))\right)^2}{4\alpha_2^2\kappa^2}+
\frac{\omega^2}{4\alpha_2\kappa^2}\frac{1}{\cosh^2(\frac{1}{2}\omega(t-C_1)}.
\end{equation}

From the equation (\ref{Phi_b}) we deduce the following equality:
$$\tanh\left(\frac{1}{4}\omega\left(t-C_1\right)\right)=
\tan\left(-\frac{\kappa\sqrt{-\alpha_2}}{2\sqrt{2}}\left(\phi-C_2\right)\right)$$
and using the expression of $\tanh\left(\frac{1}{2}\omega\left(t-C_1\right)\right)$ and $\cosh\left(\frac{1}{2}\omega
\left(t-C_1\right)\right)$ 
as functions of \linebreak $\tanh\left(\frac{1}{4}\omega\left(t-C_1\right)\right)$ we,  finally, obtain  the periodic potential
\begin{equation}
\label{V_b}
\begin{array}{l}
\displaystyle
V(\phi)=-\frac{1}{8\alpha_2^2\kappa^2}\left[-6\alpha_1^2-3\omega^2-\alpha_2\omega^2
+(3-\alpha_2)\omega^2\cos(\sqrt{2}\sqrt{-\alpha_2}\kappa(\phi-C_2))\right.\\
\displaystyle
\hskip3.5truecm
\left.+12\alpha_1\omega\sin\left(\frac{\sqrt{-\alpha_2}\kappa(\phi-C_2)}{\sqrt{2}}\right)\right].
\end{array}
\end{equation}
For the first model  (\ref{V_a}), it was not possible for us to choose the parameters $\alpha_1,\alpha_2,\alpha_3$ 
in order to have a solution (\ref{H_a}), (\ref{Phi_a})  that is an attractor for the dynamics of the system and at the same time to obtain a spectral index $n_s$ near of the experimental value  $0.96$.
The same conclusion was obtained in ref.~\cite{HMAASJY} for the particular case $\alpha_1=0$.

In the following we will consider only the second model (\ref{V_b}) whose parameters can be chosen in such a way that the solution 
(\ref{H_b}), (\ref{Phi_b}) is an attractor of the dynamics and  agrees with the recent observational data. It is important to note that the particular solution (\ref{H_b}), (\ref{Phi_b}) is the only solution of the system defined by the potential (\ref{V_b}) that satisfies the condition (\ref{eqe}). The three arbitrary parameters of this system will be chosen in such a way that this particular explicit solution be an attractor of the dynamics.

\section{The periodic potential}

Let us consider the second model defined by the potential (\ref{V_b}) with the particular exact solution  (\ref{H_b}), (\ref{Phi_b}).
 
As previously mentioned, in this case we must have $\alpha_2<0$. When $\alpha_1\neq0$ we can take  $\alpha_3>0$ and have $\omega$ real.
If $\alpha_1=0$ (which is the constant-roll condition) it is necessary to take $\alpha_3<0$. As  $\alpha_2<0$ we introduce the positive parameter $\beta$ as follows: 
$$\alpha_2=-\beta^2.$$
Note that $C_2$ is a parameter of the system and not an arbitrary integration constant because this parameter appears in the expression of the potential.

In the following, without loss of generality, we take the integration constant $C_1=0$. Also, we fix the parameter $C_2$ in such way 
that $\lim_{ {t\to -\infty }}\phi(t)=0 $, i.e.
\begin{equation}
\label{C_2}
C_2=\frac{\pi}{\sqrt{2}\beta\kappa}.
\end{equation} 

The particular exact solution (\ref{H_b}), (\ref{Phi_b}) is now written as
\begin{equation}
\label{H1}
H(t)=-\frac{\omega\tanh(\frac{1}{2}\omega t)-\alpha_1}{2\beta^2},
\end{equation} 
\begin{equation}
\label{phi1}
\phi(t)=\frac{\pi}{\sqrt{2}\beta\kappa}+\frac{2\sqrt{2}\arctan[\tanh(\frac{1}{4}\omega t)]}{\beta \kappa},
\end{equation}
where
$$\omega=\sqrt{\alpha_1^2-4\beta^2\alpha_3}.$$

From the expression of $H(t)$ we obtain the scale factor:

\begin{equation}
\label{a}
a(t)=a_0 e^{\frac{\alpha_1 t}{2 \beta ^2}} \cosh ^{-\frac{1}{\beta ^2}}
\left(\frac{\omega t }{2}\right),
\end{equation} 
where $a_0$ is an arbitrary  positive integration constant. We see from this expression that the scale factor does not vanish for any finite value of $t$.

In order to have $H(t)>0$ we must take $\alpha_1>0$ and $\alpha_3>0$ . It easy to see that in this case we have
$\lim_{ {t\to \infty }}a(t)=\infty$  and when $\alpha_3<0$ 
we have $\lim_{ {t\to \infty }}a(t)=0$. 
When $\alpha_3=0$  the limit of the scale factor is a constant value.

The potential (\ref{V_b}) can be rewritten in a simple form: 
\begin{equation}
\label{V1}
V(\phi)= F \cos (\Omega  \phi )+G \cos (2 \Omega  \phi )+E,
\end{equation} 
where 
\begin{equation}
\label{EFG}
    E= \frac{6 \alpha_1^2-\beta ^2 \omega ^2+3 \omega ^2}{8 \beta ^4 \kappa ^2}, \hskip0.5truecm
		F= \frac{12\alpha_1 \omega }{8 \beta ^4 \kappa ^2}, \hskip0.5truecm G=\frac{\left(\beta ^2+3\right) \omega ^2}{8 \beta ^4 \kappa ^2} , \hskip0.5truecm \Omega= \frac{\beta  \kappa }{\sqrt{2}} .
\end{equation} 

This potential is a periodic function of $\phi$ where the period is  
\begin{equation}
\label{periodphi}
T = \frac{2\pi}{\Omega} = \frac{2 \sqrt{2} \pi }{\beta  \kappa }.
\end{equation}

Let us examine the extrema of $V(\phi)$ within a period. These extrema are given by the zeros of the first derivative:
\begin{equation}
\label{deriphi}
V'(\phi)=-\Omega  \sin (\Omega  \phi ) (F+4 G \cos (\Omega  \phi )).
\end{equation}
The factor $\sin (\Omega  \phi )$ vanishes (within a period) for 
$$\phi=0, \hskip0.5truecm \phi=\frac{\pi}{\Omega}, \hskip0.5truecm \phi=\frac{2\pi}{\Omega}. $$

The other factor $F+4 G \cos (\Omega  \phi )$ vanishes for:
$$\phi=\frac{1}{\Omega}\arccos\left(-\frac{3 \alpha_1}{\left(\beta ^2+3\right) \omega }\right),  \hskip0.5truecm
\phi=\frac{2\pi}{\Omega}-\frac{1}{\Omega}\arccos\left(-\frac{3 \alpha_1}{\left(\beta ^2+3\right) \omega }\right).$$
These two values are real if
$\frac{3 \alpha_1}{\left(\beta ^2+3\right) \omega }<1$.
In this case the potential has the form of a  double well.

When 
$\frac{3 \alpha_1}{\left(\beta ^2+3\right) \omega }=1$,
the double well disappears.

If 
\begin{equation}
\label{positV}
\frac{3 \alpha_1}{\left(\beta ^2+3\right) \omega }\geq1
\end{equation}
$V(\phi)$ has a single well with a minimum at the half period $\phi=\frac{\pi}{\Omega}$. 
It is  easy to see that this minimum is positive:
$$V(\frac{\pi}{\Omega})=E-F+G=\frac{3 (\alpha_1-\omega )^2}{4 \beta ^4 \kappa ^2}> 0 ,$$

hence $V(\phi)$ is positive. For our particular solution~(\ref{phi1}), the potential takes values only in a half period with a maximum at $\phi(-\infty)=0$ and a minimum at 
$\phi(+\infty)=\frac{ \sqrt{2} \pi }{\beta  \kappa }=\frac{\pi}{\Omega}$.

Let us examine the  stability of our particular exact solution (\ref{H1}), (\ref{phi1}).
The equations of motion  (\ref{eqa}),  (\ref{eqb}),  (\ref{eqc}) are not independent. We can deduce  equation (\ref{eqb}) from (\ref{eqa}) and (\ref{eqc}).

This allows us to study the stability of the solution (\ref{H1}), (\ref{phi1}) in terms of  a two-dimensional dynamical system in the phase space defined by 

$$x_1=\phi(t),  \hskip0.5truecm  x_2=\dot{\phi}.$$

Since $H(t)>0$ the equation (\ref{eqa}) can be replaced by

$$H=\sqrt{\frac{\kappa^2}{3}(\frac{1}{2}x_2^2+V(x_1))},$$

and the equation (\ref{eqc}), combined with the preceding expression of $H$, can be replaced by the two dimensional autonomous system:

$$\dot{x}_1=x_2, \hskip0.5truecm \dot{x}_2=-3\sqrt{\frac{\kappa^2}{3}(\frac{1}{2}x_2^2+V(x_1))}x_2-V'(x_1), $$

which contains all the dynamics of the system. The critical points in a period are given by $x_2=0$ and $V'(x_1)=0$, that is:

$$ (0, 0),\hskip0.5truecm \left(\frac{\pi}{\Omega}, 0\right).$$ 

The first point is reached by  the exact particular solution(\ref{phi1}) at $t\to -\infty$ and the second point by  the (same) exact particular solution at $t\to +\infty$. This means that the exact particular solution is a separatrix for the dynamics, also known as an instanton in other contexts. 

It is easy to verify that the first critical point is a saddle point. Therefore this point is not an attractor of the dynamics and 
we are only interested in the second critical point.

The eigenvalues associated with the second critical point are:
\begin{equation}
\label{eigen}
\left(-\frac{\omega}{2}, \hskip0.5truecm \frac{-3\alpha_1+(\beta^2+3)\omega}{2\beta^2}\right).
\end{equation}
If $\frac{3\alpha_1}{(\beta^2+3)\omega}>1$, these eigenvalues are strictly negative, this critical point is hyperbolic, it is a stable node and then the particular solution~(\ref{phi1}) is an attractor.

If $\frac{3\alpha_1}{(\beta^2+3)\omega}=1$, the second eigenvalue vanishes, hence this critical point is no longer hyperbolic and it is not a stable node. It can be shown that the solution is not an attractor in this case.

In the following section we will choose the parameters $\alpha_1,\beta,\alpha_3$ so that 
$\frac{3 \alpha_1}{\left(\beta ^2+3\right) \omega }>1$ since  the exact solution  must  be an attractor of the dynamics. In fact, the parameters will be chosen in such a way that this quantity will be  very slightly above 1.
As we have a two dimensional dynamical system, any solution can flow  to a critical point,  a periodic orbit, or escape to infinity. 

However, this system doesn't admit periodic solutions, as can be, easily, deduced from eq.~(\ref{eqb}), because the left hand side is a derivative ant the rigth hand side is of definite sign. 

Moreover, when $3\alpha_1/((3+\beta^2)\omega)>1$, there exists  only one attractive critical point, within any given period of the potential.

Therefore, any given solution can only, either escape to infinity, or approach the critical point~$(\frac{\pi}{\Omega}, 0)$.

If a solution does escape to infinity, it is very easy to deduce how it does so, because the potential is bounded. The asymptotic behavior, obtained from the equations of motion~(\ref{eqa}),~(\ref{eqb}) and~(\ref{eqc}), can be found to be
\begin{equation}
\label{toinfty}
H(t)\sim\frac{1}{3t-K_1},\,\hskip0.5truecm \phi(t)\sim\frac{\sqrt{6}}{3\kappa}\ln\left(3t-K_1\right)+K_0,
\end{equation}
where $K_0$ and $K_1$ are arbitrary constants. This behavior implies that the solution escapes to infinity in finite time. 

For this class of solutions, the scale factor behaves as 
\begin{equation}
\label{scalefactorinfty}
a(t)\sim a_0 (3t-K_1)^{1/3},
\end{equation}
where $a_0$ is an arbitrary constant. 

Therefore, the solutions that are relevant for physics, flow to the critical point~$(\frac{\pi}{\Omega}, 0)$, as $t\to+\infty$. 
It is in this sense that we can state that our particular  solution~(\ref{H1}), ~(\ref{phi1}), is an attractor, since all physically relevant solutions will behave as it does, for large values of time. 
We end this section by giving some quantities useful in the following.  

The slow-roll parameters $\epsilon_1(t)=-\frac{\dot{H}(t)}{H^2(t)}$ ,   
$\epsilon_2(t)=\frac{\dot{\epsilon}_1(t)}{H(t)\epsilon_1(t)}$ ,   
$\epsilon_3(t)=\frac{\dot{\epsilon}_2(t)}{H(t)\epsilon_2(t)}$ ,  are given by:

\begin{equation}
\label{epsilon1}
\epsilon_1(t)=\frac{\beta ^2 \omega ^2}{\left(\alpha_1 \cosh \left(\frac{\omega t }{2}
\right)-\omega  \sinh \left(\frac{\omega t}{2}\right)\right)^2} ,
\end{equation}

\begin{equation}
\label{epsilon2}
\epsilon_2(t)= \frac{2 \beta ^2 \omega  \left(\omega -\alpha_1 \tanh 
\left(\frac{ \omega t}{2}\right)\right)}{\left(\alpha_1-\omega \tanh 
\left(\frac{\omega t}{2}\right)\right)^2} ,        
\end{equation}

\begin{equation}
\label{epsilon3}
\epsilon_3(t)= \frac{\beta^2 \omega \text {sech}^2\left(\frac{\omega t}{2}\right) \left(\alpha_1^2+\alpha_1 \omega  \tanh \left(\frac{\omega t}{2}\right)-2 \omega ^2\right)}{\left(\alpha_1 \tanh \left(\frac{\omega t}{2}\right)-\omega\right) 
\left(\alpha_1-\omega \tanh \left(\frac{\omega t}{2}\right)\right)^2} .
\end{equation}

The function $\ddot{a}(t)$ has two zeros given by:

\begin{equation}
\label{t1}
 t_1=\frac{1}{\omega}\ln\left[\frac{\left( \beta^4\left(\alpha_1^4+16\beta^2\alpha_3^2\left(1+\beta^2\right)-4\alpha_1^2\alpha_3\left(1+2\beta^2\right)\right)\right)^{1/2}+2\alpha_3\beta^2+4\alpha_3\beta^4-\alpha_1^2\beta^2}{\alpha_1\omega-\alpha_1^2+2\alpha_3\beta^2  } \right],
\end{equation}

\begin{equation}
\label{t2}
t_2=\frac{1}{\omega}\ln\left[\frac{\left( \beta^4\left(\alpha_1^4+16\beta^2\alpha_3^2\left(1+\beta^2\right)-4\alpha_1^2\alpha_3\left(1+2\beta^2\right)\right)\right)^{1/2}-2\alpha_3\beta^2-4\alpha_3\beta^4+\alpha_1^2\beta^2}{-\alpha_1\omega+\alpha_1^2-2\alpha_3\beta^2  } \right].
\end{equation}
When the parameters are chosen in such a way that these two quantities are real, $t_1$ represents the end of the inflation and $t_2$ the beginning of a new period of acceleration.

The conformal time $\tau=\int\frac{dt}{a(t)}$ can be expressed in terms of the Gauss hypergeometric functions:

$$\tau(t)=-\frac{2^{1-\frac{1}{\beta ^2}}
   \beta ^2 e^{-\frac{\alpha_1 t}{2 \beta ^2}} \left(e^{\omega t}+1\right)^{-\frac{1}{\beta ^2}} \left(e^{-\frac{\omega t}{2}} \left(e^{\omega t}+1\right)\right)^{\frac{1}{\beta ^2}} \, _2F_1\left(-\frac{1}{\beta ^2},-\frac{\alpha_1+\omega }{2 \beta ^2 \omega },1
	-\frac{\alpha_1+\omega }{2 \beta ^2 \omega },-e^{\omega t}\right)}{\alpha_1+\omega }. $$
The integration constant has been fixed in order to have $\tau(\infty)=0$.

We define the time $t^*$ by the condition that it is at 65 e-folds before the time $t_1$ of  the end of inflation,  {\em viz.}
\begin{equation}
\label{tetoile}
\ln\left(\frac{a(t_1)}{a(t^*)}\right)-65=0
\end{equation}
and $t^{**}$  by the condition that it is at 10 e-folds after the time $t^*$, {\em viz.}
\begin{equation}
\label{t2etoiles}
\ln\left(\frac{a(t^{**})}{a(t^*)}\right)-10=0.
\end{equation}
In this way, between $t^*$ and $t_1$ we have 65 e-folds and between $t^*$ and $t^{**}$  we have 10 e-folds. We will employ these two quantities later.

\section{Scalar  perturbations}

We consider the gauge invariant curvature perturbations $\zeta_k.$ In order to study these scalar perturbations and determine the spectral index $n_s$ we must analyze the Mukhanov-Sasaki (M-S) equation~\cite{mukhanov,sasaki} 
for the mode function $v_k=\sqrt{2}M_{Pl}z\zeta_k$:
\begin{equation}
\label{mu-sa}
v_k''+\left(k^2-\frac{z''}{z}\right)v_k	=0 ,
\end{equation}
where a prime denotes a derivative with respect to the conformal time $\tau$ and $z=a \sqrt{\epsilon_1}.$
The potential term $\frac{z''}{z}$ is exactly expressed in terms of the slow-roll parameters as follows: 
\begin{equation}
\label{pot Z}
 \frac{z''}{z}= a^2(t)H^2(t)W(t)  ,   %(24)
\end{equation}
 where
\begin{equation}
\label{W}
W(t)=-\frac{1}{2}\epsilon_1(t) \epsilon_2(t)-
\epsilon_1(t)+\frac{1}{2} \epsilon_2(t) \epsilon_3(t)+
\frac{\epsilon_2(t)^2}{4}+\frac{3 \epsilon_2(t)}{2}+2 . 
\end{equation}
Starting from the sub-Hubble regime where $k^2>>\frac{z''}{z}$, the M-S equation reduces to $v_k''+k^2v_k	=0$. 
We choose the adiabatic vacuum boundary condition, i.e no particles at $\tau\rightarrow-\infty$:
\begin{equation}
\label{adiabatic}
v_k \approx\frac{\exp{(-i k\tau)}}{\sqrt{2k}}.            
\end{equation}

In order to solve the M-S equation we shall approximate the potential term $\frac{z''}{z}$.
The factor $W(t)$ is replaced by its value at $t^*$. Here we do {\em not} employ the usual slow--roll approximation.
As we will see later, $W(t)$  will be almost practically constant from 
$t\approx-\infty$ to $t\approx t^{**}$.
 We introduce also the usual approximation 
$a H \approx -\frac{1}{\tau}$ which is justified in a quasi-de Sitter situation where 
$H\approx$ constant. 

So, we can approximate the potential term by a simple expression dependent  of a parameter $\nu$ defined as follows:    
\begin{equation}
\label{eqnu}
\frac{z''}{z}\approx \frac{1}{\tau^2}W(t^*)=\frac{\nu^2-1/4}{\tau^2}.
\end{equation}
For determining $\nu$ we choose the positive root of the equation (\ref{eqnu}).
 
With these approximations  equation (\ref{mu-sa}) becomes
\begin{equation}
\label{eqhankel}
v_k''+\left(k^2-\frac{\nu^2-1/4}{\tau^2}\right)v_k	=0 .
\end{equation}
With  (\ref{adiabatic}) as boundary condition, the solution of  (\ref{eqhankel}) is given
by
\begin{equation}
\label{soleqhankel}
v_k=\frac{\sqrt{-\pi\tau}}{2}H_\nu^{(1)}(-k\tau),
\end{equation}
where $H_\nu^{(1)}$ is the  Hankel function of index $\nu$.

From this explicit solution we obtain the power spectrum of the curvature perturbation:
\begin{equation}
\label{powspect}
\Delta_s^2(k)=\frac{k^3}{2\pi^2}\left|\zeta_k\right|^2=
\frac{H^2}{8\pi^2 M_{Pl}^2 \epsilon_1}\left(\frac{k}{a H}\right)^3\frac{\pi}{2}
\left|H_\nu^{(1)}(-k\tau)\right|^2.
\end{equation}
Using the asymptotic expression, when $x<<1$, for $H_\nu^{(1)}(x)$,
$$H_\nu^{(1)}(x) \approx -\frac{i}{\pi}\Gamma(\nu)\left(\frac{x}{2}\right)^{-\nu},$$ 
we obtain, in the super horizon limit  $-k \tau = \frac{k}{a H}<<1$:
\begin{equation}
\label{powspect1}
\Delta_s^2(k)= \frac{H^2}{8\pi^2 M_{Pl}^2 \epsilon_1}\frac{2^{2\nu-1}\Gamma^2(\nu)}{\pi}\left(\frac{k}{a H}\right)^{3-2\nu}.
\end{equation}
The spectral index $n_s$ is given by
\begin{equation}
\label{specindex}
n_s-1 = \frac{d\ln(\Delta_s^2(k))}{d\ln(k)} = 3-2\nu.
\end{equation}

\section{Determination of the parameters}

We will impose that our model  give the experimental value~\cite{planck} 
$$n_s=0.96.$$

We shall show that it is possible to choose the values of the remaining parameters in such a way that we are above the bifurcation value where the double well disappears. In this way the particular solution will be an attractor of the dynamics.

With this aim we introduce a positive parameter $\epsilon$ as follows: 

$$\frac{3 \alpha_1}{\left(\beta ^2+3\right) \omega }=1+\epsilon.$$
 
In fact we shall see that $\epsilon$ must be a ``small number''. When $\epsilon$ increases from zero up to a critical value $\epsilon^*$, the value of $t_1$ tends to $t_2$, i.e. $t_1(\epsilon^\ast)=t_2(\epsilon^\ast)$. The quantity $t_2-t_1$ is a decreasing function of $\epsilon$. When $\epsilon\to\epsilon^\ast$, the period of time without accelerated expansion tends to zero. In our model  we have $\ddot{a}(t)>0$ in the intervals $(-\infty , t_1)$ 
and $(t_2 , +\infty)$ .
From the preceding equation and the definition of $\omega$ given in section 2 we obtain:

$$\alpha_1= \frac{\omega\left(\beta^2+3\right)(1+\epsilon)}{3},$$

$$\alpha_3=\frac{(\beta^2+3\epsilon+\beta^2\epsilon)(6+\beta^2+3\epsilon+\beta^2\epsilon)\omega^2}{36\beta^2}.$$
Here $\beta$ and $\epsilon$ are dimensionless parameters. We will see in the following that all the relevant physical quantities depend on $\omega$ in a very simple way.
From these expressions and the results given in section 3 we deduce:

$$\ln\left(a(t)\right)=\omega t(\frac{1}{6}+\frac{1}{2\beta^2}+\frac{\epsilon}{6}+\frac{\epsilon}{2\beta^2})-\frac{\ln\left( \cosh(\frac{\omega t }{2})\right)}{\beta^2}+\ln(a_0),$$
 
$$\epsilon_1(t)=\frac{\beta ^2}{\left(\frac{1}{3}(3+\beta^2)(1+\epsilon)\cosh \left(\frac{\omega t }{2}
\right)-\sinh \left(\frac{\omega t}{2}\right)\right)^2},$$

$$\epsilon_2(t)= \frac{2 \beta ^2 \left(1 -\frac{1}{3}\left(\beta^2+3\right)(1+\epsilon) \tanh 
\left(\frac{ \omega t}{2}\right)\right)}{\left(\frac{1}{3}\left(\beta^2+3\right)(1+\epsilon)-\tanh 
\left(\frac{\omega t}{2}\right)\right)^2} ,$$

$$\epsilon_3(t)= \frac{\beta^2 \text {sech}^2\left(\frac{\omega t}{2}\right) \left((\frac{1}{3}\left(\beta^2+3\right)(1+\epsilon))^2+ \frac{1}{3}\left(\beta^2+3\right)(1+\epsilon) \tanh \left(\frac{\omega t}{2}\right)-2 \right)}{\left(\frac{1}{3}\left(\beta^2+3\right)(1+\epsilon)\tanh \left(\frac{\omega t}{2}\right)-1\right) 
\left(\frac{1}{3}\left(\beta^2+3\right)(1+\epsilon)-\tanh \left(\frac{\omega t}{2}\right)\right)^2} ,$$
\begin{equation}
\label{newt1}
\begin{array}{l}
\displaystyle
t_1=\frac{1}{\omega}\ln\left\{
-\frac{ \beta^4(1+\epsilon)^2+9\epsilon(2+\epsilon) +6\beta^2(-2+\epsilon(2+\epsilon))}{\left(3\epsilon+\beta^2(1+\epsilon)\right)^2} \right.\\
\displaystyle
\left.\hskip2.2truecm
-\frac{6\beta\left[  
-\beta^4(1+\epsilon)^2-9\epsilon(2+\epsilon)-3\beta^2(-1+2\epsilon(2+\epsilon))
\right]^{1/2}
}{\left(3\epsilon+\beta^2(1+\epsilon)\right)^2}
\right\}.
\end{array}
\end{equation}
As we can see, the parameter $\omega$ only appears multiplied by $t$ in the expressions of $\ln[a(t)]$, $\epsilon_1(t)$, $\epsilon_2(t)$, and $\epsilon_3(t)$. We see also that the product 
$\omega t_1$ is independent of the parameter $\omega$.

The quantity $\epsilon^*$ is determined by the condition $t_2-t_1=0$, which can be reduced, after some simple calculations,  to the following equation:

$$1+\beta^2-\frac{1}{9}(3+\beta^2)^2(1+\epsilon)^2=0 .$$

The only root that is not always negative is 
$$\epsilon^*=\frac{-9-6\beta^2-\beta^4+3\sqrt{9+15\beta^2+7\beta^4+\beta^6}}{9+6\beta^2+\beta^4}.$$

This quantity is positive for $0<\beta<\sqrt{3}$ with a maximum value $\frac{3}{2\sqrt{2}}-1$ $\approx 0.06066$  for $\beta=1$.

Then, for a given value of $\beta$,  $\epsilon^*$ is fixed and we must have $0<\epsilon<\epsilon^*$.

Equations (\ref{tetoile}) and (\ref{eqnu}) becomes:

$$\ln\left(\frac{a(t_1)}{a(t^*)}\right)-65=0,$$

$$-\frac{1}{2}\epsilon_1(t^*) \epsilon_2(t^*)-\epsilon_1(t^*)+\frac{1}{2} \epsilon_2(t^*)\epsilon_3(t^*)+\frac{\epsilon_2(t^*)^2}{4}+\frac{3 \epsilon_2(t^*)}{2}+2 =\nu^2-1/4,$$

where  $\nu=1.52$ is determined from the equation (\ref{specindex}) with $n_s=0.96$ and  
$t_1$ is given by (\ref{newt1}).
From these two equations we can determine $t^*$ and one of the parameters of the model, i.e. $\omega, \epsilon$, or $\beta$. As we have said above, it is easy to see from the expressions 
of $\epsilon_1(t), \epsilon_2(t), \epsilon_3(t), ln[a(t)]$ and $t_1$ that in the system of equations $\omega$ appears only multiplied by $t^*$. Let us remark the important fact 
that  the quantity $a(t_1)$ is independent of $\omega$. In consequence, it is not possible to fix in an arbitrary way the values of the parameters $\beta$ and $\epsilon$. We can fix the parameter $\beta$, for instance, and determine the values of $\epsilon$ and the product $\omega t^*$. 
If we fix $\beta$ between $0$ and $\sqrt{3}$ in order to obtain a positive value of $\epsilon$, we obtain a real solution of the system of two equations with $\epsilon>0$ only in a narrow interval 
of $\beta$ given approximately by (0.1978740, 0.1981544).

If we take $\beta=0.1981544$
the numerical solution of the  system of two equations is:
$$\epsilon=0.00619158,$$ 
$$t^*=-2.41005/\omega.$$
For these values we obtain:
$$t_1=4.51615/\omega,$$  
$$t_2=4.7784/\omega.$$
These values are relatively close which means that we are near of $\epsilon^*=0.006273064$, which is the maximum value of $\epsilon$ for the corresponding value of $\beta=0.1981544$ that gives $t_1=t_2=4.64306$.
 
If we take
$$\beta=0.1978740$$
the numerical solution is:
$$\epsilon=2.5125\times10^{-12},$$ 
$$t^*=-2.44519/\omega.$$
For these values we obtain
$$t_1=3.73286/\omega,$$  
$$t_2=6.34417/\omega.$$
Since $\epsilon\approx10^{-12}$ is close to zero, $\beta$ is close to the minimum.

To summarize ,we have approximately:
$  0.1978740 < \beta < 0.1981544$    
which gives
$ 0.00619158 >\epsilon>0$

We can now justify the approximations to $\frac{z''}{z}$. We have verified for several values of the parameters that the function $W(t)$  is slowly varying  in the interval $(-\infty,t^{**})$. 
In order to give an example of this behaviour we take $\beta=0.197874068$ . The corresponding value of $\epsilon$ is $10^{-8}$. If 
we take for instance $\omega=1.31898\times10^{-17}\mathrm{s^{-1}}$ which gives $H(\infty)=H_0=2.1983\times10^{-18}\mathrm{s^{-1}}$,  i.e. the current value of the Hubble parameter,
 we obtain $t_1=2.83\times10^{17}\mathrm{s}$, $t_2=4.8\times10^{17}\mathrm{s}$, $t^*=-1.85\times10^{17}\mathrm{s}$ and $t^{**}=-1.52322\times10^{17}\mathrm{s}$. From these values we obtain $W(-\infty)=2.05871$, $W(t^*)=2.06041$, $W(t^{**})=2.06131$ and 
$$\frac{W(t^{**})-W(t^*)}{W(t^*)}=0.00043833.$$
The function $H(t)$ is slowly decreasing in the interval $(-\infty,t^{**})$ : 
$H(-\infty)=3.39066\times10^{-16}\mathrm{s}^{-1}$, $H(t^*)=3.12062\times10^{-16}\mathrm{s}^{-1}$, $H(t^{**})=2.99232\times10^{-16}\mathrm{s}^{-1},$ hence:
$$\frac{H(t^{*})-H(t^{**})}{H(t^*)}=0.041.$$
 This allows us to focus on  the generation of perturbations of interest not only in the interval  $(t^*,t^{**})$ but also in the  
interval $(-\infty, t^{**})$~\cite{HMAAS}.

We end this section by expressing  the potential in terms of the parameters chosen above: 
$$
\begin{array}{l}
\displaystyle
V(\phi)=\frac{\omega^2}{2\beta^4\kappa^2}
\left(
(3+\beta^2)(1+\epsilon)\cos\left(\Omega\phi\right) + 
\frac{3+\beta^2}{4}\cos\left(2\Omega\phi\right)+\right.\\
\displaystyle
\hskip2.7truecm
\left.
\frac{1}{12}\left(27 + 9\beta^2+2\beta^4+36\epsilon+24\beta^2\epsilon+4\beta^4\epsilon +18\epsilon^2+12\beta^2\epsilon^2+2\beta^4\epsilon^2\right)
\right),
\end{array}
$$
where $\Omega = \beta\kappa/\sqrt{2}.$

Obviously, by changing the value of the energy scale factor $\omega$ that is arbitrary in our model we can change the order of magnitude of the multiplicative coefficient of the potential as we want.

\section{ Tensor perturbations}

We consider the tensor perturbations $\delta g_{ij}=a^2h_{ij}.$ Here $h_{ij}$ has two polarizations 
$h_{,\lambda}$  where $\lambda = +\, \mathrm{and}\, \times.$

In order to study these tensor perturbations and determine the power spectrum we must analyze the equation for the mode function $u_{k,\lambda}=\frac{a}{2} M_{Pl}h_{k,\lambda}$:
\begin{equation}
\label{equklambda1}
u_{k,\lambda}''+\left(k^2-\frac{a''}{a}\right)u_{k,\lambda}	= 0 , 
\end{equation}
where a prime denotes a derivative with respect to the conformal time. The potential term 
$\frac{a''}{a}$ is given by
$$\frac{a''}{a}= a^2 H^2(2-\epsilon_1).$$   
We employ the same type of approximation used before for the scalar perturbations. With these approximations equation~(\ref{equklambda1}) becomes:
\begin{equation}
\label{equklambda2}
u_{k,\lambda}''
+\left(k^2-\frac{1}{\tau^2}\left(2-\epsilon_{1}(t^*)\right)\right)u_{k,\lambda}	= 0 . 
\end{equation}
We introduce the parameter $\nu_t$ as follows
$$\nu_t^2-\frac{1}{4}=2-\epsilon_{1}(t^*),$$ 
i.e.
$$\nu_t=\sqrt{9/4-\epsilon_1(t^*)}.$$
The solution of the evolution equation (\ref{equklambda2}), with the vacuum boundary condition 
$u_{k,\lambda}(\tau)\approx \exp(-i k \tau)/\sqrt{2k}$ 
when $\tau\rightarrow-\infty$, is given by
$$u_{k,\lambda}=\frac{\sqrt{-\pi \tau}}{2}H^{(1)}_{\nu_t}(-k\tau).$$

From this solution we obtain the power spectrum of $ h_{k,\lambda}$ 

$$\Delta^2_{t,\lambda}(k)=\frac{k^3}{2\pi^2}\left|h_{k,\lambda}\right|^2
=\frac{H^2}{2\pi M_{Pl}^2}\left(\frac{k}{a H}\right)^3\left|H^{(1)}_{\nu_t}(-k\tau)\right|^2.$$
Since we have two polarizations, we obtain
$$\Delta^2_{t}(k)= 2\Delta^2_{t,\lambda}(k)
= \frac{H^2}{\pi M_{Pl}^2}\left(\frac{k}{a H}\right)^3\left|H^{(1)}_{\nu_t}(-k\tau)\right|^2.$$
Using the asymptotic behaviour of the Hankel function $H^{(1)}_{\nu_t}(x)$ for $x<<1$, we deduce: 
$$\Delta^2_{t}(k)=\frac{H^2}{\pi^3 M_{Pl}^2}2^{2\nu_t}
\Gamma^2(\nu_t)\left(\frac{k}{a H}\right)^{3-2\nu_t}.$$
The tensor to scalar ratio $r$ is, therefore,  given by the expression:
$$r=\frac{\Delta^2_{t}(k)}{\Delta_s^2(k)}=8 \epsilon_1(t^*) \frac{2^{2\nu_t}}{2^{2\nu-1}} 
\frac{\Gamma^2(\nu_t)}{\Gamma^2(\nu)} \left(\frac{k}{a H}\right)^{2(\nu-\nu_t)}.$$
Here $r$ is evaluated at $k =a H$ with $\nu = 1.52$ , $\nu_t=\sqrt{9/4-\epsilon_1(t^*)}=1.4989$.
The obtained values of $r$ are monotonously increasing with $\epsilon$ and, of course, independent of $\omega$. For $\epsilon=0$ we have
$r=0.05191$ and for $\epsilon=0.00619$ we obtain $r=0.05339$.
These values  are compatible with the constraint imposed by the latest measurements~\cite{planck}.

\section{Super-Hubble evolution}

In this section we will show that the curvature perturbations $\zeta_k$ become constant in the super-Hubble regime.

In the super-Hubble regime $k^2<<\frac{z''}{z}$ the evolution equation (\ref{mu-sa}) becomes:
\begin{equation}
\label{mu-sa-sup-hub}
v_k''-\frac{z''}{z} v_k	=0 .
\end{equation}
The solution of (\ref{mu-sa-sup-hub}) is
\begin{equation}
\label{vk-sup-hub}
v_k	= A_k z + B_k z\int\frac{d\tau}{z^2}, 
\end{equation}
where $A_k$ and $B_k$ are integration constants. We deduce the curvature perturbations
\begin{equation}
\label{dzetak-sup-hub}
\zeta_k	= A_k + B_k\int\frac{dt}{a^3\epsilon_1}, 
\end{equation}
with
\begin{equation}
\label{solu-integr}
\begin{array}{l}
\int\frac{dt}{a(t)^3\epsilon_1(t)}= I(t).
\end{array}
\end{equation}
The function $I(t)$ is increasing, since the integrand is positive.   Here $I(t)$ takes  negative values between $t^*$ and $t_2$ as we see below, hence its absolute value 
is decreasing. Choosing $a(0)=1$ and taking  for instance the values chosen above $\beta=0.197874068$ and $\omega= 1.31898\times10^{-17}\mathrm{s^{-1}}$ we obtain:
$$I(t^*)=-8.67311\times10^{78} \mathrm{s},$$
$$I(t^{**})=-5.50356\times10^{65} \mathrm{s},$$
$$I(0)=-5.0561\times10^{16} \mathrm{s},$$
$$I(t_1)=-4.105\times10^{-7} \mathrm{s},$$
$$I(t_2)=-1.032\times10^{-7} \mathrm{s}.$$
In summary, in the super-Hubble regime, the two modes of the curvature perturbations are  a constant 
mode and a decaying mode. Only the constant mode survives which is, in fact, necessary to transmit the information from inflation.
 
\section{Conclusions}

In this paper we have generalized a model proposed in references~\cite{JMHMTS} and~\cite{HMAASJY}. We have introduced a constraint in the form of a differential equation,   which generalizes the  constant roll constraint.

While the physical meaning of this  constraint, for the moment, isn't obvious, the consequences for inflationary models can be straightforwardly obtained:  It does lead to  an exact and explicit particular solution of an inflationary model with a simple periodic potential,  that depends on  three  parameters $\alpha_1$, $\alpha_2$, $\alpha_3$. It should be stressed that  this solution is an attractor of the dynamics. 
    
If these parameters satisfy the inequalities $\frac{3 \alpha_1}{\left(\beta ^2+3\right) \omega }>1$, $\alpha_2<0$,
 $\alpha_1>0$, $\alpha_3>0$, i.e. $\epsilon>0$, $\beta$ real and $\omega>0$, the periodic potential $V(\phi)$ is positive with a single well within any given  period and the evolution of the Hubble parameter $H(t)$ and the scalar field $\phi(t)$ has the following properties: 
 
\begin{itemize}
\item There is no initial singularity in finite co-moving time. 
\item The scalar field $\phi(t)$ begins at $t=-\infty$  on the maximum 
of the potential  $V(\phi)$, the 
field $\phi(t)$ increases during half a period of $V(\phi)$ until $t=+\infty$ where the minimum of $V(\phi)$ is reached.
\item The scale factor $a(t)$ describes  inflation from $t=-\infty$, where it vanishes, to $t_1$, a deceleration from
 $t_1$ to $t_2$ and a new acceleration from $t_2$ to  $t=+\infty$.
 \item This behavior is, indeed, typical of any solution of physical relevance, because all such solutions have been shown to behave in the same way, for large times.
 \end{itemize}
 We have imposed to our model the experimental value $n_s=0.96$ for the spectral index.
As we are only interested in the inflationary  period, we have limited the study of the model up to $t_1$. 
The parameter $\beta$ can only take values in the interval $0.1978740 <\beta < 0.1981544$ , approximately, in order to obtain a real solution for $t^*$ an a real positive solution for $\epsilon$. 
The corresponding interval of variation of $\epsilon$ is found to be $(0, \, 0.00619158)$, approximately. 
The deduced interval for the tensor to scalar ratio is $(0.05191, \, 0.05339)$, compatible with the constraint imposed by the latest experimental results~\cite{planck}.
The parameter $\omega$ remains in our model a free positive parameter that can fix an adequate scale energy for the potential and the time.             

\vskip0.2truecm

{\bf Acknowledgements}: We would like to warmly thank  David Polarski for very useful discussions about inflationary models and  Stam Nicolis for a careful and critical reading of the manuscript.

\end{document}